\documentclass[
showkeys,12pt,
preprint,preprintnumbers,nofootinbib,
groupedaddress,superscriptaddress,amsmath,amssymb]{revtex4}
\usepackage{graphicx}
\usepackage{dcolumn}
\usepackage{bm}
\usepackage{amssymb}
\usepackage{amsmath}
\usepackage{epsfig}    
\usepackage{color}
\usepackage{slashed}
\usepackage{hhline}

\def\be{\begin{equation}}
\def\ee{\end{equation}}
\newcommand{\bea}{\begin{eqnarray}}
\newcommand{\eea}{\end{eqnarray}}
\newcommand{\nn}{\nonumber}

\numberwithin{equation}{section}

\begin{document}

{\begin{flushright}{KIAS-P16035}
\end{flushright}}

\title{An $SU(2)_L$ Septet Scalar Linking to Radiative Neutrino Model}
%
%
\author{Takaaki Nomura}
\email{nomura@kias.re.kr}
\affiliation{School of Physics, KIAS, Seoul 130-722, Korea}

\author{Hiroshi Okada}
\email{macokada3hiroshi@gmail.com}
\affiliation{Physics Division, National Center for Theoretical Sciences, Hsinchu, Taiwan 300}

\author{Yuta Orikasa}
\email{orikasa@kias.re.kr}
\affiliation{School of Physics, KIAS, Seoul 130-722, Korea}
\affiliation{Department of Physics and Astronomy, Seoul National University, Seoul 151-742, Korea}

\date{\today}

\begin{abstract}
We study a radiative neutrino model with a SU(2) septet scalar with hypercharge $Y=2$ where active neutrino mass is induced at a one-loop level.
Calculating a one-loop diagram, mass matrices for active neutrino is derived. 
We also analyze lepton flavor violations, anomalous magnetic moment of muon and the flavor violating Higgs decay $h \to \mu \tau$.
Then sizable muon $g-2$ and $\sim 1 \%$ of $BR(h \to \mu \tau)$ could be obtained in our set up.
Furthermore, the collider signature of the septet is discussed where we focus on quadratically charged scalar at the LHC 14 TeV.
We find that it could be discovered up to $1.5$ TeV mass by searching multi charged lepton signal events. 

\end{abstract}
\maketitle
\newpage

\section{Introduction}

After the discovery of the Higgs boson which is consistent with Standard Model(SM)~\cite{Chatrchyan:2012xdj, Aad:2012tfa}, we have been working for further understanding of the Higgs sector.
 The structure of Higgs sector is still uncertain and there would be other scalar contents besides SM Higgs.
 We expect that new scalar fields are connected to issues which can not be explained by the SM such as neutrino masses and mixing, dark matter(DM), Dark energy, matter anti-matter asymmetry and so on.
 Actually many models beyond the SM introduce new scalar fields extending the Higgs sector.
  
  The radiative seesaw models are typical cases which extend the Higgs sector.
  In these models active neutrino masses and mixing are generated at loop level with new TeV scale particles where they often include DM candidate.
  Thus these models could be one of  the natural realization of the tiny neutrino mass at TeV energy scale involving DM,  which would be verifiable by various experiments in the near future.  
  Along this line of ideas,  a vast amount of papers has been arisen in Refs~\cite{Zee, Cheng-Li, zee-babu,Pilaftsis:1991ug, Krauss:2002px, Ma:2006km, Aoki:2008av, Gustafsson:2012vj, Hambye:2006zn, Gu:2007ug, Sahu:2008aw, Gu:2008zf, Babu:2002uu, AristizabalSierra:2006ri, AristizabalSierra:2006gb,
Nebot:2007bc, Bouchand:2012dx, Kajiyama:2013sza,McDonald:2013hsa, Ma:2014cfa, Schmidt:2014zoa, Herrero-Garcia:2014hfa,
Ahriche:2014xra,Long1, Long2, Aoki:2010ib, Kanemura:2011vm, Lindner:2011it,
Kanemura:2011jj, Aoki:2011he, Kanemura:2011mw, Schmidt:2012yg, Kanemura:2012rj, Farzan:2012sa, Kumericki:2012bf, Kumericki:2012bh, Ma:2012if, Gil:2012ya, Okada:2012np, Hehn:2012kz, Baek:2012ub, Dev:2012sg, Kajiyama:2012xg, Kohda:2012sr, Aoki:2013gzs, Kajiyama:2013zla, Kajiyama:2013rla, Kanemura:2013qva,Law:2013saa, Dasgupta:2013cwa,Toma:2013zsa, Baek:2013fsa, Baek:2014qwa, Okada:2014vla, Ahriche:2014cda, Ahriche:2014oda,Chen:2014ska,
Kanemura:2014rpa, Okada:2014oda, Fraser:2014yha, Okada:2014qsa,Vicente:2014wga, Hatanaka:2014tba,Sierra:2014kua, Baek:2015mna, Jin:2015cla,
Culjak:2015qja, Okada:2015nga, Geng:2015sza, Okada:2015bxa, Geng:2015coa, Ahriche:2015wha, Restrepo:2015ura, Kashiwase:2015pra, Nishiwaki:2015iqa, Wang:2015saa, Okada:2015hia, Ahriche:2015loa, Ahn:2012cg, Ma:2012ez, Kajiyama:2013lja, Hernandez:2013dta, Ma:2014eka, Aoki:2014cja, Ma:2014yka, Ma:2015pma, Ma:2013mga,
radlepton1, radlepton2, Okada:2014nsa, Brdar:2013iea, Okada:2015nca, 
Okada:2015kkj, Fraser:2015mhb, Fraser:2015zed, Adhikari:2015woo, Kanemura:2015cca, Bonnet:2012kz,Sierra:2014rxa, Davoudiasl:2014pya, Lindner:2014oea,Okada:2014nea, MarchRussell:2009aq, King:2014uha, Mambrini:2015sia, Boucenna:2014zba, Ahriche:2016acx, Okada:2015vwh, Nomura:2016fzs, Nomura:2016seu, Ahriche:2016rgf, Nomura:2016run, Nomura:2016vxr,Sierra:2016qfa, Lu:2016ucn,Merle:2016scw, Kownacki:2016hpm, Thuc:2016qva, Ahriche:2016cio, Ahriche:2016ixu, Ma:2016nnn,Rocha-Moran:2016enp}. 
  In radiative seesaw models, SU(2) multiplet scalars are often introduced such as doublet and triplet, in order to achieve observed neutrino mass and mixing.
Among the SU(2) multiplet scalar higher than doublet, the septet scalar $\Phi_7$ with hypercharge $Y=2$ is particularly interesting since vacuum expectation value(VEV) of it retain $\rho =1$ at the tree level~\cite{Hisano:2013sn, Alvarado:2014jva, Geng:2014oea,Hamada:2015bra,Aranda:2015xoa}. 
Thus it is interesting to consider a radiative seesaw model which includes the septet whose VEV is connected in generating neutrino masses.

In this paper, we consider a radiative seesaw model with the septet scalar in which the neutrino mass is generated at the one-loop level.
We note that an accidental global U(1) is conserved in the Higgs sector of the doublet and the septet which would induce massless Nambu-Goldstone boson (GB) when the septet develops a VEV~\cite{Hisano:2013sn, Alvarado:2014jva, Geng:2014oea,Hamada:2015bra,Aranda:2015xoa}.
This issue {in our model} can be solved by adding SU(2) triplet scalars and a signet scalar inducing non-renormalizable operator that breaks the global U(1) symmetry.
These new scalars also play a role in generating active neutrino mass. 
Then we derive the active neutrino mass matrix analyzing a loop diagram involving the septet.
We also discuss lepton flavor violations(LFVs) and muon anomalous magnetic moment $(g-2)_\mu$ where  sizable $(g-2)_\mu$ could be obtained satisfying constraints from LFVs. 
In addition flavor violating Higgs decay $h \to \tau \mu$ is also analyzed where
the CMS reported an excess of the events with significance of 2.4$\sigma$ and the best fit 
value of branching ratio is $BR(h \to \mu \tau) = (0.84^{+0.39}_{-0.37})\%$~\cite{Khachatryan:2015kon}; ATLAS'  best fit value is $BR(h \to \mu \tau) = (0.77\pm 0.62)\%$~\cite{Aad:2015gha},  consistent with but less significant than CMS.
Furthermore, collider signature of the septet would be interesting since it contain multi-charged scalar bosons as its components.
We perform some simulation study for the multi-lepton signal events and discuss the discovery potential for multi-charged component of the septet.


This paper is organized as follows.
In Sec.~II, we show our model,  including neutrino sector, LFVs, muon anomalous magnetic moment and $h \to \tau \mu$.
In Sec.~III, we perform a simulation study for the signature of the septet.
We conclude and discuss in Sec.~IV.


\section{ Model setup}
 \begin{widetext}
\begin{center} 
\begin{table}
\begin{tabular}{|c||c|c|c|c||c|c|c|c|c|c|}\hline\hline  
&\multicolumn{4}{c||}{Lepton Fields} & \multicolumn{5}{c|}{Scalar Fields} \\\hline
& ~$L_L$~ & ~$e_R^{}$~ & ~$L'$~ & ~$N_R$ ~ & ~$\Phi$ ~ & ~$\Phi_7$~  & ~$\Delta_1$~  & ~$\Delta_0$~ & ~$S_0$ \\\hline 
$SU(2)_L$ & $\bm{2}$  & $\bm{1}$  & $\bm{2}$ & $\bm{1}$ & $\bm{2}$ & $\bm{7}$ & $\bm{3}$& $\bm{3}$ & $\bm{1}$  \\\hline 
$U(1)_Y$ & $-\frac12$ & $-1$  & $-\frac{1}{2}$& $0$  & $\frac12$ & ${2}$ & $1$  & $0$ & $0$  \\\hline
 $Z_2$ & $+$ & $+$   & $-$ & $-$& $+$  & $+$ & $-$& $+$  & $-$  \\\hline
\end{tabular}
\caption{Contents of fermion and scalar fields
and their charge assignments under $SU(2)_L\times U(1)_Y\times Z_2$.}
\label{tab:1}
\end{table}
\end{center}
\end{widetext}

In this section, we explain our model. 
The particle contents and their charges are shown in Tab.~\ref{tab:1}.
We add {three (or two)} iso-spin doublet vector-like exotic fermions $L'$ with $-1/2$ hypercharge, an isospin singlet scalar $S_0$, two isospin triplet scalars $\Delta_i$  with $i=0,1$ hypercharge, and an isospin septet scalar $\Phi_7$  with $2$ hypercharge to the SM, where 
$S_0$ can be a real field for brevity, while the neutral component of $\Delta_1$ has to be a complex field.
Therefore the neutrino masses are induced not  through the $S_0$ mediation but through the $\Delta_1$ one.
We assume that neutral components of $\Phi$, $\Delta_0$, and $\Phi_7$ have vacuum
expectation values (VEVs), which are respectively symbolized by $v/\sqrt2$ and $v_\Delta/\sqrt2$ and $v_7/\sqrt2$. 

The relevant Lagrangian and Higgs potential under these symmetries are given by
\begin{align}
-\mathcal{L}_{Y}
&=
(y_{\ell})_{ij} \bar L_{Li} \Phi e_{Rj} +(y_{\Delta})_{ij} \bar L^c_{L_i}(i\tau_2) \Delta_1 L'_{L_j} 
 + (y_N)_{ij} \bar L'_{L_i}\tilde\Phi N_{R_j} \nn\\
 & + (y_L)_{ij} \bar L_{L_i} L'_{R_j} S_{0} +  (M_{L})_{i} \bar L'_{L_i} L'_{R_i}+  (M_{R})_i \bar N^c_{R_i} N_{R_i} + {\rm h.c.}, \\
\mathcal{V}&=  \mathcal{V}_{\rm nontrivial\ term} + \mathcal{V}_2+\mathcal{V}_4,\\
 \mathcal{V}_{\rm nontrivial\ term}&=\lambda_0 \Delta_1 \Phi_7^* \Delta_1\Delta_0+\mu_0\Phi^\dag \Delta_0\Phi+\mu_1\Phi^T(i\tau_2)\Delta_1^\dag\Phi S_0 +{\rm c.c.},
\label{Eq:lag-flavor}
\end{align}
where $i=1-3$, $j=1-3$, $\tau_i(i=1-3)$ is Pauli matrix, $\tilde\Phi\equiv i\tau_2\Phi^*$, $\mathcal{V}_2$ is the trivial quadratic term, $\mathcal{V}_4$ is the trivial quartic term, and the first term of $\mathcal{L}_{Y}$ generates the SM
charged-lepton masses $m_\ell\equiv y_\ell v/\sqrt2$ after the electroweak spontaneous breaking of $\Phi$.
We work on the basis where all the coefficients are real and positive for simplicity. 
The scalar fields can be parameterized as 
\begin{align}
&\Phi =\left[
\begin{array}{c}
w^+\\
\frac{v+h+iz}{\sqrt2}
\end{array}\right],\quad 
\Delta_1 =\left[
\begin{array}{cc}
\frac{\Delta^+}{\sqrt2} & \Delta^{+2}\\
\Delta^{0} & -\frac{\Delta^+}{\sqrt2}
\end{array}\right],
\quad 
\Delta_0 =\left[
\begin{array}{cc}
\frac{\delta^0}{\sqrt2} & \delta^{+}\\
\delta^{-} & -\frac{\delta^0}{\sqrt2},
\end{array}\right],\\
&\Phi_7= \left[\varphi^{3},\varphi^{2},\varphi^{1},\varphi^{0},\varphi^{-1},\varphi^{-2},\varphi^{-3} \right]^T =
\left[\phi^{+5},\phi^{+4},\phi^{+3},\phi^{+2},\phi^{+}_2,\phi^{0},\phi^{-}_1\right]^T,
\label{component}
\end{align}
where $\Delta^0\equiv \frac{\Delta_R+i\Delta_I}{\sqrt2}$, $\delta^0\equiv \frac{v_\Delta+\delta_R+i\delta_I}{\sqrt2}$, $\phi^0\equiv \frac{v_7+\phi_R+i\phi_I}{\sqrt2}$, $v_{\rm SM}=\sqrt{v^2+v_7^2+v_\Delta^2}\simeq 246$ GeV is VEV of the SM-like  Higgs, and index for the components in $\Phi_7$ denotes the electric charge for $\phi$ and the eigenvalue of $T_3$ for $\varphi$.
{\it Here we assume to analyzed that all the mixings are negligible for simplicity}, although
each of component mixes as follows:
The CP-even mass matrix with nonzero VEVs is written in terms of the basis $[h,\delta_R^0,\phi_R^0]$ basis. 
The CP-odd mass matrix with nonzero VEVs is written in terms of the basis $[z,\delta_I^0,\phi^0_I]$ basis, but 
the lightest state is massless that is absorbed by the longitudinal component of $Z$ boson. Thus the mass matrix is reduced to be  $2\times2$. 
The inert CP-even mass matrix  is written in terms of $[S_0,\Delta_R]$ basis.
The singly charged mass matrix with nonzero VEVs is written in terms $[w^\pm,\delta^\pm,\phi_1^\pm,\phi_2^\pm]$ basis, but 
the lightest state is massless that is absorbed by the longitudinal component of $W^\pm$ boson. Thus the mass matrix is reduced to be  $3\times3$. 
Here  the other fields [$\Delta_I$, $\Delta^{\pm}$,  $\Delta^{\pm2}$, $\phi^{\pm2}$,   $\phi^{\pm3}$, $\phi^{\pm4}$, $\phi^{\pm5}$] do not mix each other.      

We also comment that the non-renormalizable operator $\Phi_7 \Phi^5 \Phi^*$ is obtained by triplet scalars and singlet scalar effects at one-loop level, which is required to break an accidental global U(1) symmetry 
in the Higgs sector with the doublet and the septet. Thus our model does not induce massless Nambu-Goldstone boson when $\Phi_7$ develops the VEV.

\subsection{ Neutral component of fermions}
In our model, we have two neutral fermions $N_R$ and $N'$, where $N$ is  the neutral component of $L'$ and
we define $L'\equiv [N',E']^T$ here. These fields mix through the term $y_N$, where we assume $y_N$ to be diagonal and $M_N\equiv M_L\approx M_R$.
Then we simply parametrize the mixing matrix as follows:
\begin{align} 
&\left[
\begin{array}{c}
N'^c_L\\
N_R
\end{array}\right]
\approx
\left[
\begin{array}{cc}
\frac1{\sqrt2} &-\frac1{\sqrt2}\\
\frac1{\sqrt2} & \frac1{\sqrt2}
\end{array}\right]
\left[
\begin{array}{c}
N^c_1\\
N_2
\end{array}\right],\\
&M_1= 
M_N-m_D,
\quad
M_2=
M_N+m_D,
\end{align}
where $m_D\equiv y_Nv/\sqrt2$, and each of $M_{\sigma_i} (\sigma=1,2)$, $i$ is the flavor index,  and $[N^c_1,N_2]^T$ is the mass eigenvalue and eigenstate.

\subsection{ Neutrino mass matrix}
When $S_0$ and $\Delta_0$ are complex fields, we have two sources to generate the neutrino masses at the one-loop level. But since we assume  $S_0$ to be real, the neutrino masses can only be generated through the $\Delta_0$
mediation as mentioned above.
\begin{figure}[tb]\begin{center}
\includegraphics[scale=0.25]{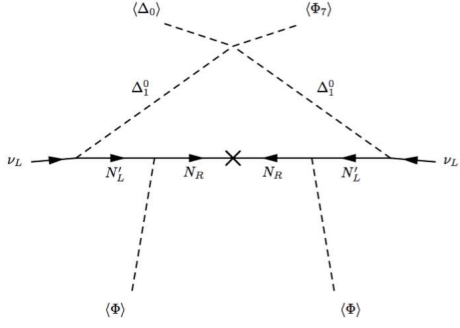}
\caption{ Neutrino masses at one-loop level.
}   \label{fig:neut1}\end{center}\end{figure}
Then the formula of active neutrino mass matrix $m_\nu$ as shown in Figure~\ref{fig:neut1} is given by 
\begin{align}
&(m_{\nu})_{ab}
=-\sum_{\sigma=1}^2\sum_{i=1}^3\frac{(y_\Delta)_{ai} M_{\sigma_i}(y_\Delta^T)_{ib} }{2(4\pi)^2}
\left[
\frac{X_{R,\sigma_i} }{X_{R,\sigma_i}-1}\ln[X_{R,\sigma_i}] - \frac{X_{I,\sigma_i} }{X_{I,\sigma_i}-1}\ln[X_{I,\sigma_i}]
 \right],
\end{align}
where we define $X_{\alpha,\sigma_i}\equiv (m_{\Delta_a}/M_{\sigma_i})^2$ with $\alpha$=R/I and $\sigma=1,2$.
VEVs of $\Delta_0$ and $\Phi_7$ play the role in generating the mass difference between $\Delta_R$ and $\Delta_I$ through the term of $\lambda_0$. 
 $({m}_\nu)_{ab}$ can be generally diagonalized by the Pontecorvo-Maki-Nakagawa-Sakata mixing matrix $V_{\rm MNS}$ (PMNS)~\cite{Maki:1962mu} as
\begin{align}
({m}_\nu)_{ab} &=(V_{\rm MNS} D_\nu V_{\rm MNS}^T)_{ab},\quad D_\nu\equiv (m_{\nu_1},m_{\nu_2},m_{\nu_3}),
\\
V_{\rm MNS}&=
\left[\begin{array}{ccc} {c_{13}}c_{12} &c_{13}s_{12} & s_{13} e^{-i\delta}\\
 -c_{23}s_{12}-s_{23}s_{13}c_{12}e^{i\delta} & c_{23}c_{12}-s_{23}s_{13}s_{12}e^{i\delta} & s_{23}c_{13}\\
  s_{23}s_{12}-c_{23}s_{13}c_{12}e^{i\delta} & -s_{23}c_{12}-c_{23}s_{13}s_{12}e^{i\delta} & c_{23}c_{13}\\
  \end{array}
\right],
\end{align}
where we neglect the Majorana phase as well as Dirac phase $\delta$ in the numerical analysis for simplicity.
The following neutrino oscillation data at 95\% confidence level~\cite{pdf} is given as
\begin{eqnarray}
&& 0.2911 \leq s_{12}^2 \leq 0.3161, \; 
 0.5262 \leq s_{23}^2 \leq 0.5485, \;
 0.0223 \leq s_{13}^2 \leq 0.0246,  
  \\
&& 
  \ |m_{\nu_3}^2- m_{\nu_2}^2| =(2.44\pm0.06) \times10^{-3} \ {\rm eV}^2,  \; 
  \ m_{\nu_2}^2- m_{\nu_1}^2 =(7.53\pm0.18) \times10^{-5} \ {\rm eV}^2, \nn
  \label{eq:neut-exp}
  \end{eqnarray}
where we assume one of three neutrino masses is zero with normal ordering in our analysis below.
%
The observed PMNS matrix can be realized by introducing the following parametrization.
Here we can parametrize the Yukawa coupling $y_L$ as follows;
\begin{align}
y_\Delta
&= V_{\rm MNS} \sqrt{D_\nu} OR^{-1/2},
\label{yl-sol}
\\
R_{ii}&\equiv
 -\sum_{\sigma=1}^2\frac{ M_{\sigma_i} }{2(4\pi)^2}
\left[
\frac{X_{R,\sigma_i} }{X_{R,\sigma_i}-1}\ln[X_{R,\sigma_i}] - \frac{X_{I,\sigma_i} }{X_{I,\sigma_i}-1}\ln[X_{I,\sigma_i}]
 \right],
 \label{R-sol}
\end{align}
where $O$ is an arbitrary complex orthogonal matrix.
Here we estimate the order of Yukawa coupling $y_\Delta$. 
First of all, we assume to be $V_{\rm MNS}= O={\cal O}$(1), and $R_{ii}=\frac{M_{\sigma_i}}{2(4\pi)^2}\times {\cal O}(10^{-2})$. Then the Eq.(\ref{yl-sol}) is simplified as 
\begin{align}
y_\Delta= {\cal O}(100)\times \sqrt{\frac{D_\nu}{M_{\sigma_i}}}\approx {\cal O}(10^{-4}),
\label{eq:Yukawa-order}
\end{align}
where we fix to be $D_\nu={\cal O}$(0.1) eV and $M_{\sigma_i}={\cal O}$(100) GeV.

\begin{table}[t]
\begin{tabular}{c|c|c} \hline
Process & $(a,b)$ & Experimental bounds ($90\%$ CL) \\ \hline
$\mu^{-} \to e^{-} \gamma$ & $(1,2)$ &
	$\text{BR}(\mu \to e\gamma) < 5.7 \times 10^{-13}$  \\
$\tau^{-} \to e^{-} \gamma$ & $(1,3)$ &
	$\text{BR}(\tau \to e\gamma) < 3.3 \times 10^{-8}$ \\
$\tau^{-} \to \mu^{-} \gamma$ & $(2,3)$ &
	$\text{BR}(\tau \to \mu\gamma) < 4.4 \times 10^{-8}$  \\ \hline
\end{tabular}
\caption{Summary of $\ell_b \to \ell_a \gamma$ process and the lower bound of experimental data~\cite{Adam:2013mnn}.}
\label{tab:Cif}
\end{table}

\subsection{ Lepton Flavor Violations, $(g-2)_\mu$, and $h\to \mu,\tau$ excess }
\label{lfv-lu}
{\it Lepton flavor violations}:
$\ell_b\to\ell_a \gamma$ processes arise from the following relevant terms at the one-loop level
\begin{align}
-\mathcal{L}_{Y}
&\supset
(y_L)_{ij}  \overline\ell_{L_i} {E'_{R_i}} S_0
+
{(y_\Delta)_{ij} } 
\overline {E'^c_{L_i}}
 \ell_{L_j} \Delta^{+2}
+
\frac{(y_\Delta)_{ij} }{2} 
(\overline {N^c_{1}} - \overline {N_{2}})_i
\ell_{L_j} \Delta^{+}.
\label{Eq:lag-mass}
\end{align}
Then the branching ratio of ${\rm BR}(\ell_b\to\ell_a \gamma)$ is defined by
\begin{align}
{\rm BR}(\ell_b\to\ell_a \gamma)
=
\frac{48\pi^3 \alpha_{em} C_b}{(4\pi)^4{\rm G_F^2} }|A_{ab}|^2,
\end{align}
where $\alpha_{em} \simeq 1/137$ is the fine structure constant, $C_i=(1,1/5)$ for ($i=e,\tau$), ${\rm G_F}\approx1.17\times 10^{-5}$ GeV$^{-2}$ is the Fermi constant, and $A_{ab}$ is  given by
\begin{align}
A_{ab}&=
(y_L^\dag)_{ai}(y_L)_{ib} F(S_0,E'_i)\nn\\
&\hspace{0cm}
-(y_\Delta^\dag)_{ai}(y_\Delta)_{ib} \left[2F(E'_i,\Delta^{+2})+F(\Delta^{+2},E'_i)+\frac{F(N_{1_i},\Delta^+)}4+\frac{F(N_{2_i},\Delta^+)}4\right]
,\\
F(c,d)&\equiv \frac{2 m_c^6+3m_c^4m_d^2-6m_c^2m_d^4+m_d^6+12m_c^4m_d^2\ln\left[\frac{m_d}{m_c}\right]}{12(m_c^2-m_d^2)^4}.
\end{align} 
Assuming here the small $y_\Delta$ in Eq.~(\ref{eq:Yukawa-order}), $A_{ab}$ can be written in terms of $y_L$ only.
Thus the resultant branching ratio is given by
\begin{align}
{\rm BR}(\ell_b\to\ell_a \gamma)
\approx
\frac{3 C_b |(y_L^\dag)_{ai}(y_L)_{ib}  F(S_0,E'_i)|^2}{16\pi^2{\rm G_F^2}}.
\end{align}
Comparing the experimental values in table.~\ref{tab:Cif}, we obtain the following upper bounds
\begin{align}
\text{BR}(\mu \to e\gamma) &:\quad
 |(y_L^\dag)_{1i}(y_L)_{i2}  F(S_0,E'_i)|\lesssim \frac{1.60\times 10^{-11}}{{\rm GeV}^2},\\
 \text{BR}(\tau \to e\gamma) &:\quad
 |(y_L^\dag)_{1i}(y_L)_{i3}  F(S_0,E'_i)|\lesssim \frac{8.59\times 10^{-9}}{{\rm GeV}^2},\\
 \text{BR}(\tau \to \mu\gamma) &:\quad
 |(y_L^\dag)_{2i}(y_L)_{i3}  F(S_0,E'_i)|\lesssim \frac{9.92\times 10^{-9}}{{\rm GeV}^2}.
\end{align}

{\it Muon anomalous magnetic moment $(g-2)_\mu$}:
Our formula of $(g-2)_\mu$ is given by
\begin{align}
\Delta a_\mu\approx 
\frac{(y_L^\dag)_{2i}(y_L)_{i2} m^2_\mu F(S_0,E'_i)}{(4\pi)^2}
,
\label{damu}
\end{align}
which can be comparable to the experimental value $20\times 10^{-10}\lesssim\Delta a_\mu^{\rm exp}\lesssim 40\times 10^{-10}$ in refs.~\cite{bennett, discrepancy1, discrepancy2}.

{\it$h\to \mu,\tau$ excess}:
The recent experiment at the LHC reports $ {\rm BR}(h\to \mu\tau)\approx$ 1\%, and 
The excess of $h\to \mu\tau$ can be generated at one-loop level as the leading contribution in our model. 
The resultant decay rate formulas are expressed as
\begin{align}
& \Gamma(h\to \mu\tau) \approx 
\frac{|(y^\dag_L)_{2i} (y_L)_{i3}\mu_{hSS}m_h m_\tau G |^2}{4096 \pi^5 m_h},\\
&G=\int\frac{\delta(x+y+z-1) y^2 dxdydz}{(z^2-z) m_\mu^2 + (x^2-y) m_\tau^2 -xz(m_h^2 - m_\mu^2 - m_\tau^2) + x M_{E'}^2+(y+z) m_{S_0}^2},
\label{eq:hmutau}
\end{align}
where $\mu_{hSS}\equiv \lambda_{\Phi S} v/2$ is the strength of the trilinear $h S_{0} S_0$ interaction.
Then the branching ratio reads
\begin{align}
 {\rm BR}(h\to \mu\tau) \approx \frac{ \Gamma(h\to \mu\tau) }{ \Gamma(h\to \mu\tau) + \Gamma(h) },
\end{align}
where $ \Gamma(h) \approx4.2\times 10^{-3}$ GeV is the total decay width of the SM Higgs boson at 125.5 GeV. 

Then we have the allowed region to satisfy the lepton flavor violations and the sizable excess of $h\to\mu\tau$,
while we obtain $(g-2)_\mu\approx (5-6)\times10^{-10}$, which is smaller than the experimental best fit value but within $3 \sigma$ uncertainty. This result is almost the same as the one obtained by ref.~\cite{Baek:2016kud}.

\section{Septet signature at the Large Hadron Collider}

\begin{figure}[t] 
\includegraphics[width=70mm]{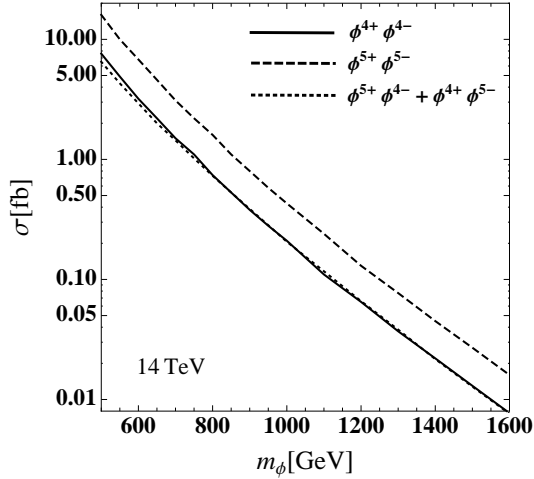} 
%
\caption{  The cross sections for the processes which provide pair of $\phi^{\pm4}$ at the LHC 14 TeV.
Here $m_{\phi^{5\pm}} = m_{\phi^{4 \pm}} + 100$ GeV is assumed for the $pp \to W^\pm \to \phi^{5 \pm} \phi^{4 \mp}$ process.
\label{fig:XS}}
\end{figure}

In this section we discuss signature of our septet scalar at the LHC.
The components of the septet can be produced via gauge interactions which come from kinetic term
\begin{equation}
(D_\mu \Phi_7)^\dagger(D^\mu \Phi_7) = (\partial_\mu \Phi_7 + i (g W_{a \mu} {\cal T}_a^{(7)} + 2 g' B_\mu ) \Phi_7)^\dagger (\partial^\mu \Phi_7 + i (g W_a^\mu {\cal T}_a^{(7)} + 2 g' B^\mu ) \Phi_7)
\end{equation}
where ${\cal T}^{(7)}_a$ are matrices for the generators of SU(2) acting on $\Phi_7^{}$. The generators ${\cal T}^{(7)}_a$ are given by 
\begin{eqnarray} & \displaystyle
{\cal T}_1^{(7)} \,\,=\,\, \frac{1}{\sqrt{2}}
\begin{pmatrix}
0 & \sqrt{3} & 0 & 0 & 0 & 0 & 0 \\
\sqrt{3} & 0 & \sqrt{5} & 0 & 0 & 0 & 0 \\
0 & \sqrt{5} & 0 & \sqrt{6} & 0 & 0 & 0 \\
0 & 0 & \sqrt{6} & 0 & \sqrt{6} & 0 & 0 \\
0 & 0 & 0 & \sqrt{6} & 0 & \sqrt{5} & 0  \\
0 & 0 & 0 & 0 & \sqrt{5} & 0 & \sqrt{3}  \\
0 & 0 & 0 & 0 & 0 & \sqrt{3} & 0
\end{pmatrix} , \nonumber \\ & \displaystyle
{\cal T}_2^{(7)} \,\,=\,\, \frac{i}{\sqrt{2} }
\begin{pmatrix}
0 & -\sqrt{3} & 0 & 0 & 0 & 0 & 0 \\
\sqrt{3} & 0 & -\sqrt{5} & 0 & 0 & 0 & 0 \\
0 & \sqrt{5} & 0 & -\sqrt{6} & 0 & 0 & 0 \\
0 & 0 & \sqrt{6} & 0 & -\sqrt{6} & 0 & 0 \\
0 & 0 & 0 & \sqrt{6} & 0 & -\sqrt{5} & 0  \\
0 & 0 & 0 & 0 & \sqrt{5} & 0 & -\sqrt{3}  \\
0 & 0 & 0 & 0 & 0 & \sqrt{3} & 0
\end{pmatrix} ,
& \nonumber \\ & \displaystyle
{\cal T}_3^{(7)} \,\,=\,\, {\rm diag}(3,2,1,0,-1,-2,-3) ~.&
\end{eqnarray}
After electroweak symmetry breaking, we obtain the kinetic term in terms of mass eigenstate of gauge fields such that  
\begin{align}
|D_\mu \Phi_7|^2 = \sum_{m=-3}^{3} \biggl| & \left[ \partial_\mu - i (2+m) e A_\mu - i \frac{g}{c_W} (m - (2+m) s_W^2) Z_\mu \right] \varphi^{m} \nonumber \\
& -i g \sqrt{ \frac{(3+m)(4-m)}{2} } W^+_\mu \varphi^{m-1} -i g \sqrt{ \frac{(3-m)(4+m)}{2} } W^-_\mu \varphi^{m+1} \biggr|^2,
\end{align}
from which we can derive relevant gauge interactions.
The decay processes of the components are induced by gauge interaction and $\lambda_0 \Delta_1 \Phi_7^* \Delta_1$ coupling in the potential Eq.~(\ref{Eq:lag-flavor}). 
The possible decay modes are 
\begin{align}
& \phi^Q \to \phi^{Q \pm1} W^\mp \\
& \phi^Q \to \Delta^{Q'} \Delta^{Q''} \quad (Q= Q'+Q''),
\end{align}
where $Q'(Q'')$ take possible charge of the components in $\Delta_1$.
Then components in $\Delta_1$ can decay through Yukawa coupling with $L'$ and $L$, and charged components in $L'$ can decay into charged lepton and $S^0$ where $S^0$ further decay into SM neutrino and DM. 
We find that the decay process of $\phi^{\pm 4}$ is particularly interesting since the decay modes of the component provide clear signal as 
\begin{equation}
\label{eq:decay4}
\phi^{\pm 4} \to \Delta^{\pm 2} \Delta^{\pm 2} \to E^\pm \ell^\pm E^\pm \ell^\pm \to \ell^\pm \ell^\pm \ell^\pm \ell^\pm S_0 S_0
\end{equation}
where $\ell^\pm$ is the SM leptons, $E^\pm$ is the exotic charged leptons and $S^0$ becomes missing transverse energy.
Thus we focus on the processes which produce pair of $\phi^{\pm4}$ in the following analysis.
The $\phi^{\pm 4}$ can be pair produced thorough the processes
\begin{align}
\label{eq:processes}
& pp \to Z/\gamma \to \phi^{4+} \phi^{4-} \nonumber \\
& pp \to Z/\gamma \to \phi^{5+} \phi^{5-} \to \phi^{4+} \phi^{4-} W^+ W^- \nonumber \\
& pp \to W^\pm \to \phi^{5 \pm} \phi^{4 \mp} \to \phi^{4+} \phi^{4-} W^\pm 
\end{align}
where $W^\pm$ in the final state can be off-shell.
In Fig.~\ref{fig:XS}, we show the cross sections for the production processes where we used  CalcHEP~\cite{Belyaev:2012qa} with {\tt CTEQ6L} PDF~\cite{Nadolsky:2008zw} in the estimation. 

We consider detector level simulation to estimate discovery potential at the LHC 14 TeV.
In order to generate the events, we employ the event generator {\tt MADGRAPH/MADEVENT\,5}~\cite{Alwall:2014hca}, where the necessary Feynman rules and relevant parameters of the model are implemented by use of FeynRules 2.0 \cite{Alloul:2013bka} and the {\tt NNPDF23LO1} PDF~\cite{Deans:2013mha} is adopted. 
Then the  {\tt PYTHIA\,6}~\cite{Ref:Pythia}  is applied to deal with hadronization effects,  the  initial-state radiation (ISR) and final-state radiation (FSR) effects, and the decays of SM particles e.g. $W$-boson, $t$-quark, etc. 
In addition, the generated events are also run though the {\tt PGS\,4} for detector level simulation~\cite{Ref:PGS}.

 \begin{widetext}
\begin{center} 
\begin{table}
\begin{tabular}{|c c c c c c|}\hline
& $n_L \geq 4$ \qquad & $n_L \geq 5$ \qquad & $n_L \geq 6$ \qquad & $n_L \geq 7$ \qquad & $n_L \geq 8$ \\ \hline
\# of signal events ($e$) & 33. & 30. & 22. & 12. & 4.2 \\ 
\# of signal events ($\mu$) & 34.  & 31.  & 25.  &  15. & 5.6  \\ \hline
\end{tabular}
\caption{Number of signal events after selecting number of leptons with $L = 100$fb$^{-1}$, $m_{\phi^{\pm4}} =1100$ GeV and $m_{\phi^{\pm5}} =1200$ GeV,
summing over events from processes in Eq.~(\ref{eq:processes}). The second and third low represent the cases where leptons from $\phi^{\pm 4}$ decay are all (anti-)electrons and (anti-)muons respectively.}
\label{tab:events}
\end{table}
\end{center}
\end{widetext}
The final state of the signal events is the multi-lepton state where
we do not put any constraint on number of jets or missing transverse energy since multi-lepton condition suppress the SM backgrounds sufficiently.
In our analysis, we assume the branching fraction of the decay chain in Eq.~(\ref{eq:decay4}) is 1 and charged leptons in the final state are either all electron or all muon type.
We also fix masses of exotic lepton and $S_0$ as $m_{L'} = 400$ GeV and $m_{S^0}= 200$ GeV for illustration; the results are not sensitive to these values as long as the decay chain is kinematically allowed.
In Table.~\ref{tab:events}, we show number of multi-lepton events for each minimum number of charged lepton $n_L$ applying luminosity of $L = 100$fb$^{-1}$, $m_{\phi^{\pm4}} =1100$ GeV and $m_{\phi^{\pm5}} =1200$ GeV
where we summed events from processes in Eq.~(\ref{eq:processes}) and final state leptons from $\phi^{\pm 4}$ decay are all electro(muon) type in second(third) low.
For selected charged leptons, basic transverse momentum cut is also applied as $p_T (\ell) > 15$ GeV.  
We find that number of remaining events become small when we require $n_L \geq 7$ although $\phi^{+4} \phi^{-4}$ provides eight charged leptons.
Also the efficiency for muon case is slightly better than that of electron.
The SM background comes from the processes of multi Z production such that $pp \to ZZ, ZZZ$ where some of $Z$
{decay leptonically}.  
We find that SM background is negligibly small when we take number of leptons as $n_L \geq 6$ while the number of signal events is not decreased much.
Thus we require the number of charged lepton as $n_L \geq 6$ for the event selection.

\begin{figure}[t] 
\includegraphics[width=70mm]{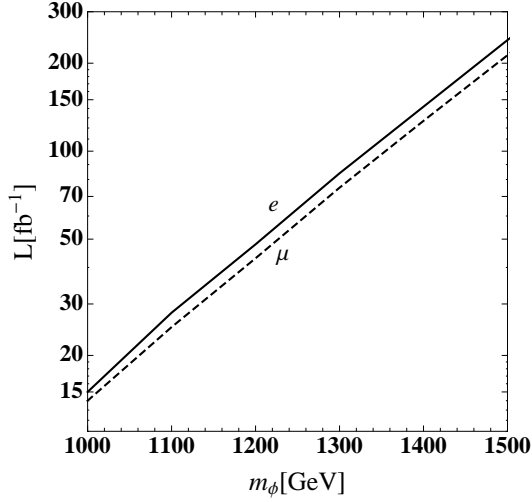} 
%
\caption{  The luminosity required to obtain 5$\sigma$ significance at the LHC 14 TeV where we took $m_{\phi^{\pm5}} =m_{\phi^{\pm4}} + 100$ GeV.
The solid(dashed) lines indicate electron(muon) dominant decay of $\phi^{\pm 4}$.
\label{fig:lumi}}
\end{figure}
\begin{figure}[t] 
\includegraphics[width=70mm]{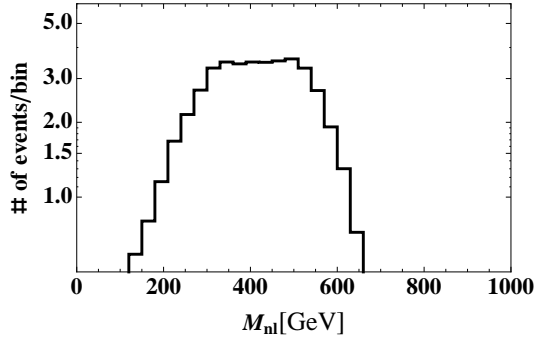} 
%
\caption{  The invariant mass for same sign leptons after the event selection where we took $L = 100$fb$^{-1}$, $m_{\phi^{\pm4}} =1100$ GeV, $m_{\phi^{\pm5}} =1200$ GeV and $m_{S^0}$ =200 GeV for illustration.
The solid and dashed lines corresponds to the cases that final state from $\phi^{\pm 4}$ decay are all electron and muon type, respectively.
\label{fig:IM}}
\end{figure}

We then estimate the resulting significance after event selection.
In our calculations,  we define the significance as~\cite{Ball:2007zza}  
\begin{equation}
S = 2 \left( \sqrt{n_s + n_b} - \sqrt{n_b} \right) \frac{n_b}{n_b + \Delta n_b^2},
\end{equation} 
where $n_s$ and $n_b$ are number of expected events for signal and background, and $\Delta n_b$ is Gaussian uncertainty of $n_b$.
Requiring $n_L \geq 6$, $n_b$ and $\Delta n_b$ are very small and we approximate the significance as $S \simeq 2 \sqrt{n_s}$ assuming $n_b \gg \Delta n_b^2$ for simplicity.
In Fig.~\ref{fig:lumi}, we show the luminosity which is required to obtain 5$\sigma$ significance at the LHC 14 TeV as a function of $\phi^{\pm 4}$ mass.
We find that $\phi^{\pm 4}$ can be discovered up to $m_{\phi^{\pm4}} \sim 1.5 $ TeV with luminosity of $L \leq 300 $fb$^{-1}$. 
Thus our model can be tested searching for multi-lepton events at the LHC.
Furthermore, distribution of invariant mass of same sign leptons is also shown in Fig.~\ref{fig:IM} where we adopt 
$L = 100$fb$^{-1}$, $m_{\phi^{\pm4}} =1100$ GeV and $m_{\phi^{\pm5}} =1200$ GeV  for illustration.
We see broad bump and the edge of the distribution around $m_{\phi^{4\pm}} - 2 m_{S^0}$.
Therefore the signature of $\phi^{4\pm}$ can be seen as a bump of invariant mass of same sign charged leptons.

\section{ Conclusions and discussions}
We have studied a one-loop induced radiative neutrino model with a SU(2) septet scalar boson in which we have discussed neutrino masses and mixing, lepton flavor violations, anomalous magnetic moment of muon, and the flavor violating SM Higgs decay $h\to\mu\tau$. 
The SU(2) triplet and singlet scalars are also introduced to break an accidental global U(1) symmetry in the Higgs sector with doublet and septet, 
which also play a role in generating active neutrino mass and mixing.
In addition, exotic lepton doublets are introduced to induce the neutrino mass at one-loop level where it also contributes to flavor violating Higgs decay. 
 
 The neutrino mass matrix has been derived calculating a one-loop diagram with the septet.
We also show that anomalous magnetic moment of muon can be within $3 \sigma$ of experimental value while satisfying the constraints from lepton flavor violating processes.
Moreover $h \to \mu \tau$ decay is indued at the one-loop level where $BR(h \to \tau \mu)$ reach $\sim 1 \%$ by tuning the relevant parameters.

Also we have explored the signature of the septet scalar by focusing on decay process of quadratically charged scalar $\phi^{\pm 4}$ which has decay chain of Eq.~(\ref{eq:decay4}). 
Since the decay provides multi-lepton final states, the signal will be very clear at collider experiments.
Then we have discussed the processes which produce pair of $\phi^{+4} \phi^{-4}$ at the LHC; $pp \to Z/\gamma \to \phi^{4+} \phi^{4-}$, $pp \to Z/\gamma \to \phi^{5+} \phi^{5-} \to \phi^{4+} \phi^{4-} W^+ W^-$ and 
$pp \to W^\pm \to \phi^{5 \pm} \phi^{4 \mp} \to \phi^{4+} \phi^{4-} W^\pm$.  
We find the production cross section is around $O(0.1)$ fb when the scale of the septet is $~1$ TeV.
To estimate the discovery potential at the LHC, we have carried out detector level simulation where the signal event is multi-lepton plus missing transverse energy.  
After simulation study, we found that $\phi^{\pm 4}$ with mass up to $\sim 1.5$ TeV can be discovered at the LHC 14 TeV for luminosity $L  \leq 300$ fb$^{-1}$.
In addition, we have shown the distribution of invariant mass for same sign charged leptons which would be broad bump.
Therefore some parameter region of our model can be tested in future LHC data.


\section*{Acknowledgments}
\vspace{0.5cm}
H. O. is sincerely grateful for all the KIAS members, Korean cordial persons, foods, culture, weather, and all the other things.
This work is supported in part by NRF Research No. 2009-0083526 (Y. O.) of the Republic of Korea.

\end{document}